\documentclass[journal]{IEEEtran}

 \usepackage{stfloats}
\usepackage{color}
\usepackage{multicol}
\usepackage[pdftex]{graphicx}
\ifCLASSINFOpdf
   \usepackage{graphicx}
\else
\fi

\usepackage[cmex10]{amsmath}
\usepackage{amssymb}
\usepackage{amsthm}
\usepackage{indentfirst}

\usepackage{algorithm}
\usepackage[noend]{algpseudocode}
\usepackage{cite}
\begin{document}

\title{A Spurious-Free Characteristic Mode Formulation Based  on Surface Integral Equation for Patch Antenna Structures}
\author{Kun~Fan,~
        Ran~Zhao,~ \IEEEmembership{Member,~IEEE,}
        Guang~Shang~Cheng,~ \IEEEmembership{Member,~IEEE,}
        Zhi~Xiang~Huang,~ \IEEEmembership{Senior Member,~IEEE,}
        Jun~Hu,~ \IEEEmembership{Senior Member,~IEEE,}
        % <-this % stops a space
\thanks{This work was supported in part by NSFC under Grant 61801002, 62031010, U20A20164, 61971001, 61871001,  61901002, NSF of Anhui Province under Grant 1808085QF183, 1908085QF258, and Grant KJ2018A0015, and in part by the State Key Laboratory of Millimeter Waves Foundation under Grant K202007.
\emph{(Corresponding authors:  Ran Zhao)}}% <-this % stops a space
\thanks{
Kun Fan, Guang Shang Cheng, Zhi Xiang Huang and Ran Zhao are with the Key Laboratory of Intelligent Computing and Signal Processing, Ministry of Education, Anhui University, China (emai: fkyct1992@163.com, ran.zhao@kaust.edu.sa, gscheng89@ahu.edu.cn
, zxhuang@ahu.edu.cn).
Jun Hu is with the School of Electronic Science and Engineering, University of Electronic Science and Technology of China (UESTC), Chengdu 611731, China (email: hujun@uestc.edu.cn).
%Hakan Bagci (and Ran Zhao also) is with the Division of Computer, Electrical, and Mathematical Science and Engineering (CEMSE), King Abdullah University of Science and Technology (KAUST), Thuwal, Saudi Arabia. (email: ran.zhao@kaust.edu.sa)

%Ping Li is with the Key Laboratory of Ministry of Education of China for Research of Design and Electromagnetic Compatibility of High Speed Electronic Systems, Shanghai Jiao Tong University, Shanghai 200240, China, and the University of Hong Kong Shenzhen Institute of Research and Innovation (HKU-SIRI), Shenzhen, China.
%

%
%Hakan Bagci is with the Division of Computer, Electrical, and Mathematical Science and Engineering King Abdullah University of Science and Technology (KAUST), Thuwal, Saudi Arabia. (e-mail: hakan.bagci@kaust.edu.sa)

}% <-this % stops a space
\thanks{}}  %Manuscript received April 19, 2005; revised December 27, 2012.,

\markboth{Journal of \LaTeX\ Class Files,~Vol.~11, No.~4, December~201x}%Journal of \LaTeX\ Class Files,~Vol.~11, No.~4, December~2012
{Shell \MakeLowercase{\textit{et al.}}: Bare Demo of IEEEtran.cls for Journals}


\maketitle

\begin{abstract}
Conventional surface integral equation (SIE)-based characteristic mode formulation for the patch antenna structure with a finite substrate is susceptible to the spurious (nonphysical) modes due to the dielectric part. To avoid the contamination of spurious modes, we propose a novel generalized eigenvalue formulation based on the electric field integral equation coupled Poggio-Miller-Chang-Harrington-Wu-Tsai (EFIE-PMCHWT) equation. In this formulation, the real and imaginary parts of the exterior integral operators are chosen to construct the finalized weighting matrices, to connect radiated power of the characteristic current. Compared with other SIE-based methods, this equation doesn't need additional post-processing since it can effectively avoid spurious modes. Numerical results compared with the volume-surface integral equation (VSIE) are investigated to validate the accuracy.
\end{abstract}

\begin{IEEEkeywords}
characteristic modes (CM), surface integral equation (SIE), patch structure, EFIE-PMCHWT, spurious-free.
\end{IEEEkeywords}

\IEEEpeerreviewmaketitle

\section{Introduction}

\IEEEPARstart{T}{he} theory of characteristic modes (TCM) has grown in popularity recently as it takes the intrinsic properties of the electromagnetic target such as structure, material, and size into account only and is independent of the excitations. Thus, the TCM can clearly explain the physical radiation mechanism from the analyzed objects.
The TCM was firstly introduced in the electromagnetic community by Garbacz \cite{1}. Then, Harrington and Mautz \cite{2} proposed the electric field integral equation (EFIE)-based TCM for perfect electric conductor (PEC) structures. Following these two methods, the magnetic field integral equation (MFIE)-based TCM was also introduced \cite{3} where no symmetric matrices are applied.

For dielectric objects, there are two  different approaches for the TCM formulations; one is starting from the volume-integral-equation(VIE) \cite{5}, another is from the surface- integral-equation(SIE) \cite{4}. Even though the VIE-based TCM formulation approach has robust solutions both for loss and lossless objects and is immune from spurious solutions, the volume discretization will always lead to many unknowns with expensive computational costs. To reduce the computational cost, the SIE-based Chang-Harrington formulation is preferred. However, as shown in \cite{6}, the  contamination of spurious modes in CH (Chang--Harrington) formulation leads to non-orthogonal far-field patterns. To address the issue, a post-processing method is proposed to remove spurious modes \cite{6}-\cite{8}.

Various methods were developed to avoid nonphysical modes contamination in recent years \cite{9}-\cite{12}, but spurious modes have not been completely removed. In \cite{13, 14}, by choosing an exterior (radiation-related) integral operator as a weighting operator of the generalized eigenvalue equation, spurious modes can be effectively avoided, and the eigenvalues have a clear physical interpretation.

Because of the particular structures of patch antennas, which are often composed of a metallic surface touched on the finite dielectric substrate, the EFIE-PMCHWT \cite{16}-\cite{18} are commonly chosen as the governing equation to model the electromagnetic property. If the CH-type formulation for the dielectrics were directly extended to analyze the CM of the patch antenna structures by using the EFIE-PMCHWT, it would also have the spurious mode contamination issue \cite{15}.The volume-surface integral equation (VSIE) based TCM is proposed for the printed patch antenna structures  \cite{19} to avoid the spurious mode issue.  However, the volume discretization of the substrate will also lead to tremendous computational costs. In this letter, a novel surface integral equation (EFIE-PMCHWT) based CM analysis method is proposed for printed patch antenna structures to reduce the computational requirements.

Inspired by \cite{14}, in this proposed method, the integral operator of EFIE-PMCHWT is split into the interior (material-related) integral operator and exterior (radiation-related) integral operator. By correctly choosing a combination of the real and imaginary parts of the exterior radiation operator as the right weighting operator of the generalized eigenvalue equation, the spurious mode contamination can be avoided, and the eigenvalues have a clear physical interpretation. The symmetrical EFIE-PMCHWT equation (sEFIE-PMCHWT) is also developed to further reduce the computational cost, with the same accuracy and higher efficiency than the non-symmetric one.

%The rest of this paper is organized as follows. In Section II, the formulation of the proposed MT-DD-SIE scheme for general composite objects is derived, expressions of the matrix system resulting from its discretization are provided, and a preconditioning technique developed to accelerate the iterative solution. In Sections III, numerical examples are provided to demonstrate the accuracy, efficiency, and applicability of the MT-DD-SIE scheme. Conclusions are summarized in Section IV

\section{TCM FORMULATIONS FOR PATCH STUCTURES}

%For convenience, we define the following trace operators in sub-domain $\Omega_{m}$: i) $\pi_{\tau}^{(m)}(\mathbf{u})$ representing the tangential components of $\mathbf{u}$ on the boundary $\partial\Omega_m$; ii) ${\pi_{\times}^{(m)}(\mathbf{u})}$ representing the twisted tangential components of $\mathbf{u}$ on the boundary $\partial\Omega_m$. $\hat{\mathbf{n}}_{m}$ is the unit normal vector pointing into sub-domain $\Omega_m$ on boundary $\partial\Omega_m$.

\begin{figure}[!htbp]
\centerline{\includegraphics[width=0.7\columnwidth,draft=false]{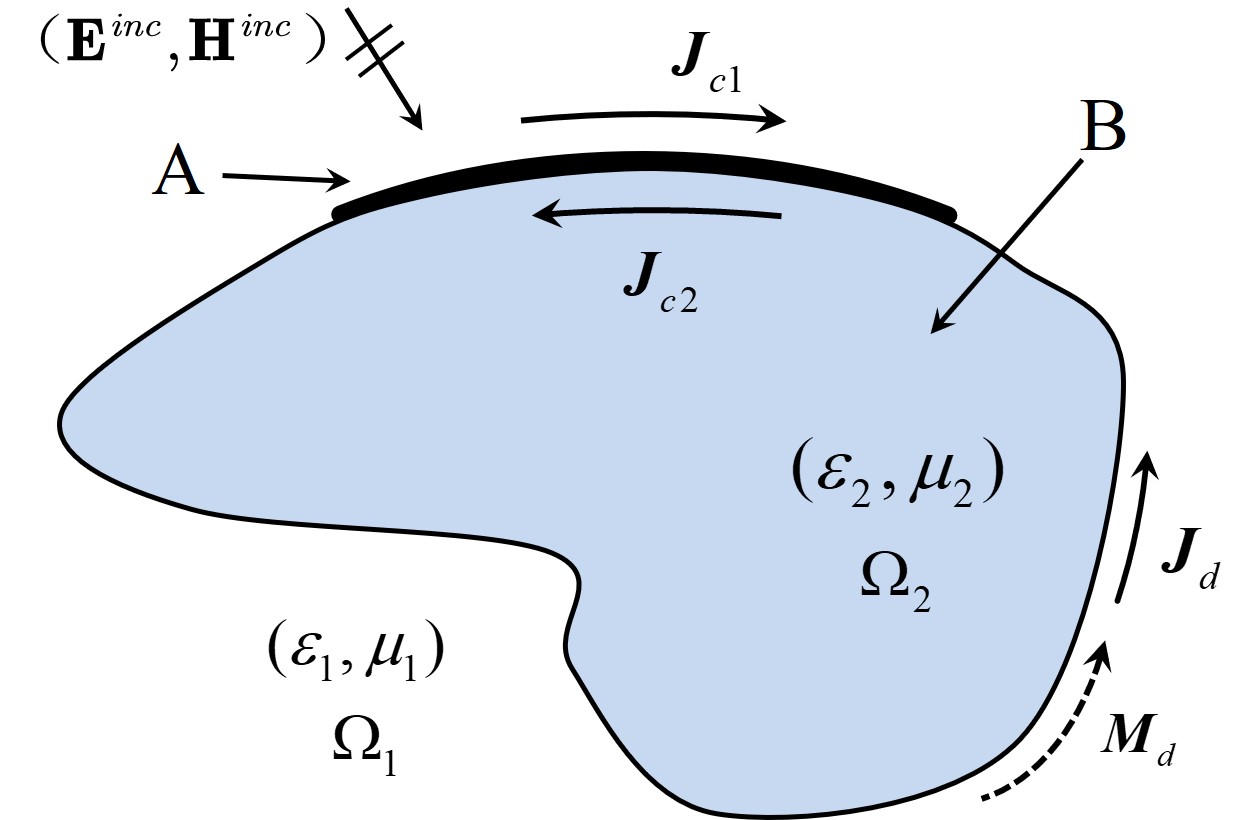}}
\caption{Electromagnetic scattering from a patch antenna structure.}
\label{fig:1}
\end{figure}

%\subsection{EFIE-PMCHWT Formulation }
As shown in Fig.~\ref{fig:1}, a simplified patch antenna structure, which consists of metallic patch A touched on the dielectric body B, is investigated. The background ${{\Omega }_{\text{1}}}$ is free space. The ${{\varepsilon }_{m}}$ and ${{\mu }_{m}}$ respectively denote permittivity and permeability of the ${{\Omega }_{m}}, m=1,2$. The ${{\eta }_{m}}\text{=}\sqrt{{{\mu }_{m}}/{{\varepsilon }_{m}}}$ are the intrinsic impedance of the region ${{\Omega }_{m}}$. Let ${{\textbf{J}}_{c1}}$and ${{\textbf{J}}_{c2}}$ represent the equivalent electric surface currents on outer and inner surface of the conducting surface and assume that ${{\textbf{J}}_{d}}$ and ${{\textbf{M}}_{d}}$ denote the equivalent electric and magnetic currents on the surface of substrate. The ${{\textbf{E}}^{\text{inc}}}$,${{\textbf{H}}^{\text{inc}}}$ denote the incident field  in exterior region.

\noindent The final linear matrix equation of the EFIE-PMCHWT formulation \cite{17, 18}  for the patch antenna structure is expressed in (1).
The subscripts $\mathrm{b,c_1,c_2}$ in this eqution denote the dielectric surface, outer and inner conducting surfaces.
\begin{figure*}[!htbp]
\begin{equation}
\left[\begin{array}{cccc}
\eta_{1} \mathbf{P}_\mathrm{d, d}^{1}+\eta_{2} \mathbf{P}_\mathrm{d, d}^{2} & -\mathbf{Q}_\mathrm{d, d}^{1}-\mathbf{Q}_\mathrm{d, d}^{2} & \eta_{1} \mathbf{P}_\mathrm{d, c_{1}}^{1} & \eta_{2} \mathbf{P}_\mathrm{d, c_{2}}^{2} \\
\mathbf{Q}_\mathrm{d, d}^{1}+\mathbf{Q}_\mathrm{d, d}^{2} & 1 / \eta_{1} \mathbf{P}_\mathrm{d, d}^{1}+1 / \eta_{2} \mathbf{P}_\mathrm{d, d}^{2} & \mathbf{Q}_\mathrm{d, c_{1}}^{1} & \mathbf{Q}_\mathrm{d, c_{2}}^{2} \\
\eta_{1} \mathbf{P}_\mathrm{c_{1}, d}^{1} & -\mathbf{Q}_\mathrm{c_{1}, d}^{1} & \eta_{1} \mathbf{P}_\mathrm{c_{1}, c_{1}}^{1} & 0 \\
\eta_{2} \mathbf{P}_\mathrm{c_{2}, d}^{2} & -\mathbf{Q}_\mathrm{c_{2}, d}^{2} & 0 & \eta_{2} \mathbf{P}_\mathrm{c_{2}, c_{2}}^{2}
\end{array}\right]\left[\begin{array}{c}
\mathbf{J}_\mathrm{d} \\
\mathbf{M}_\mathrm{d} \\
\mathbf{J}_\mathrm{c_{1}} \\
\mathbf{J}_\mathrm{c_{2}}
\end{array}\right]
=\left[\begin{array}{c}
\mathbf{b}_\mathrm{d}^\text{TE} \\
j\mathbf{b}_\mathrm{d}^\text{TH} \\
\mathbf{b}_\mathrm{c_{1}}^\text{TE} \\
0
\end{array}\right].
\label{eq:1}
\end{equation}
\end{figure*}

The detailed expression of matrix elements and vectors are listed as following,

\begin{equation}
\left\{ \begin{gathered}
\left(\mathbf{P}_{a, b}^{m}\right)_{p, q}=\left\langle \mathbf{f}_{p}^{a}, L_{m}\left(\mathbf{f}_{q}^{b}\right)\right\rangle_{a} \\
\left(\mathbf{Q}_{a, b}^{m}\right)_{p, q}=\left\langle \mathbf{f}_{p}^{a}, K_{m}\left(\mathbf{f}_{q}^{b}\right)\right\rangle_{a}^{m} \\
\left(\mathbf{b}_{a}^\text{TF}\right)_{p}=\left\langle \mathbf{f}_{p}^{a}, \mathbf{F}^{\text {inc }}\right\rangle_{a}
\end{gathered} \right..
\label{eq:3}
\end{equation}\noindent
where a and b denote the surface $\mathrm{b,c_1,c_2}$, $m=1$ or $2$, $\mathbf{F}= \mathbf{E}$ or $\mathbf{H}$ and $p$, $q$ denote the $p\mathrm{th}$, $q\mathrm{th}$ test function $\mathbf{f}_{p}^{a}$ and basis function $\mathbf{f}_{q}^{b}$ on the surface a and b.
The inner product of two vectors ${\mathbf{u},\mathbf{v}}$ are defined as
\begin{equation}
{{\left\langle \mathbf{u},\mathbf{v} \right\rangle }_{s}}=\int\limits_{S}{(\mathbf{u}\cdot \mathbf{v})dS}
\label{eq:2}
\end{equation}
\noindent The corresponding ${{\mathbf{L}}_{m}}$, ${{\mathbf{K}}_{m}}$ operators are defined as following:
\begin{equation}
\begin{split}
  & {{\mathbf{L}}_{m}}(\mathbf{X}(\mathbf{r}');\partial {{\Omega }_{m}})= \\
 & \text{      }-j{{k}_{m}}{\eta }_{m}\int\limits_{\partial {{\Omega }_{m}}}{[\mathbf{I}+\frac{1}{k_{m}^{2}}\nabla \nabla \centerdot ]{{G}_{m}}(\mathbf{r},\mathbf{r}')}\mathbf{X}(\mathbf{r}')d\mathbf{r}' \\
 & {{\mathbf{K}}_{m}}(\mathbf{X}(\mathbf{r}');\partial {{\Omega }_{m}})= \\
 & \text{      }\int\limits_{\partial {{\Omega }_{m}}}{\nabla {{G}_{m}}(\mathbf{r},\mathbf{r}')}\times \mathbf{X}(\mathbf{r}')d\mathbf{r}' \\
\end{split}
\label{eq:4}
\end{equation}
with the Green function ${{G}_{m}}(\mathbf{r},\mathbf{r}')=\frac{{{e}^{-j{{k}_{m}}|\mathbf{r}-\mathbf{r}'|}}}{4\pi |\mathbf{r}-\mathbf{r}'|}$ , and the wavenumber ${{k}_{m}}=\omega \sqrt{{{\mu }_{m}}{{\varepsilon }_{m}}}$.

The characteristic modes of the patch antenna structures can be obtained, via solving the following generalized eigenvalue equation,
\begin{equation}
\mathbf{Z}\cdot{{\mathbf{X}}_{n}}=(1+j{{\lambda }_{n}})\mathbf{W}\cdot{{\mathbf{X}}_{n}},
\label{eq:5}
\end{equation}
where $\mathbf{Z}$ is the matrix of (1), $\mathbf{W}$ is the corresponding weighting matrix, ${{\lambda }_{n}}$ and ${{\mathbf{X}}_{n}}$ are respectively the $n\mathrm{th}$ eigenvalue and corresponding eigenvector.  As shown in [14], the nonphysical modes are generated because the weighting operator $\mathbf{W}$ contains the matrix of the interior of the body, which is not related to the radiated power.

To get the proper weighing matrix without spurious mode, based on this theory, we define the exterior matrix as

\begin{equation}
{{\mathbf{Z}}^{\text{ext}}}=\left[ \begin{matrix}
   \eta_{1} \mathbf{P}_\mathrm{d, d}^{1}& -\mathbf{Q}_\mathrm{d, d}^{1}& \eta_{1} \mathbf{P}_\mathrm{d, c_{1}}^{1} & 0 \\
\mathbf{Q}_\mathrm{d, d}^{1} & 1 / \eta_{1} \mathbf{P}_\mathrm{d, d}^{1}& \mathbf{Q}_\mathrm{d, c_{1}}^{1} & 0\\
\eta_{1} \mathbf{P}_\mathrm{c_{1}, d}^{1} & -\mathbf{Q}_\mathrm{c_{1}, d}^{1} & \eta_{1} \mathbf{P}_\mathrm{c_{1}, c_{1}}^{1} & 0 \\
0 &0& 0 & 0  \\
\end{matrix} \right]
\label{eq:6}
\end{equation}
\noindent After substituting the exterior matrix (6) into the Poynting's theorem (the object is lossless),  the following equation is obtained,
\begin{equation}
\begin{split}
 & -\frac{1}{2} [\mathbf{X}_{n}^{\text{H}}\cdot({{\mathbf{Z}}^{\text{ext}}}\cdot{{\mathbf{X}}_{n}})]  = \\
 & \text{            }\frac{1}{2}\mathop{{\int\!\!\!\!\!\int}\mkern-21mu \bigcirc}\limits_S
 (\mathbf{E}\times {{\mathbf{H}}^{*}})\cdot dS +\frac{1}{2}j\omega \mathop{{\int\!\!\!\!\!\int\!\!\!\!\!\int}}\limits_V
 ({{\mu }_{1}}{{\left| \mathbf{H} \right|}^{2}}-{{\varepsilon }_{1}}{{\left| \mathbf{E} \right|}^{2}}) dV \\
\end{split}
\label{eq:7}
\end{equation}
where S denotes the surface away from the system and V denotes the space bounded by the surface S. The first term of the right-hand-side represents the radiation power and the second term represents the stored field energy. Therefore, the radiation power of the exterior part is defined as:
\begin{equation}
\mathrm{P}_{n}^{\text{rad}}=-\frac{1}{2}{\text{Re}}[ \mathbf{X}_{n}^{\text{H}}\cdot({{\mathbf{Z}}^{\text{ext}}}\cdot{{\mathbf{X}}_{n}}) ].
\label{eq:8}
\end{equation}
Inspired by [14], the weighting matrix $\mathbf{W}$ defined in \eqref{eq:13} is chosen as
\begin{equation}
\mathbf{W}\text{=}\left[ \begin{matrix}
 {\text {Re}}(\eta_{1} \mathbf{P}_\mathrm{d, d}^{1})& j{\text {Im}}(-\mathbf{Q}_\mathrm{d, d}^{1})& {\text {Re}}(\eta_{1} \mathbf{P}_\mathrm{d, c_{1}}^{1}) & 0 \\
j{\text {Im}}(\mathbf{Q}_\mathrm{d, d}^{1}) & {\text {Re}}(1 / \eta_{1} \mathbf{P}_\mathrm{d, d}^{1})& j{\text {Im}}(\mathbf{Q}_\mathrm{d, c_{1}}^{1}) & 0\\
{\text {Re}}(\eta_{1} \mathbf{P}_\mathrm{c_{1}, d}^{1}) & j{\text {Im}}(-\mathbf{Q}_\mathrm{c_{1}, d}^{1}) & {\text {Re}}(\eta_{1} \mathbf{P}_\mathrm{c_{1}, c_{1}}^{1}) & 0 \\
0 &0& 0 & 0  \\
\end{matrix} \right]
\label{eq:13}
\end{equation}
to satisfy
\begin{equation}
\mathrm{P}_{n}^{\text{rad}}=-\frac{1}{2}{\text {Re}}[\mathbf{X}_{n}^{\text{H}}\cdot({{\mathbf{Z}}^{\text{ext}}}\cdot{{\mathbf{X}}_{n}}) ] \text{=}-\frac{1}{2}[ \mathbf{X}_{n}^{\text{H}}\cdot(\mathbf{W}\cdot{{\mathbf{X}}_{n}}) ].
\label{eq:14}
\end{equation}

In the electromagnetic scattering problem, the final linear matrix equation  is defined as,
\begin{equation}
\mathbf{Z}\cdot\mathbf{X}={{\mathbf{F}}^{\text{inc}}}
\label{eq:16},
\end{equation}
where ${{\mathbf{F}}^{\text{inc}}}$ is the right-hand-side vector of the linear equation discretized from the incident field. Due to the nonsymmetry of $\mathbf{Z}$ and $\mathbf{W}$ are ,  the supplementary eigenvalue equation  [20] can be constructed as
\begin{equation}
{{\mathbf{Z}}^{\text{T}}}\cdot\mathbf{X}_{n}^{a}=(1\text{+}j{{\lambda }_{n}}){{\mathbf{W}}^{\text{T}}}\cdot\mathbf{X}_{n}^{a}
\label{eq:17},
\end{equation}
where the $\mathrm{T}$ represents the transpose of a matrix. The eigenvectors satisfy the following orthogonality.
\begin{equation}
\mathbf{X}_{m}^{\text{T}}\cdot(\mathbf{W}\cdot\mathbf{X}_{n}^{a})={{\delta }_{mn}}.
\label{eq:18}
\end{equation}
Due to the complete orthogonality of the set formed by characteristic modes  \cite{21}, the induced current generated by an external source can be defined as
\begin{equation}
\mathbf{X}\approx \sum\limits_{n=1}^{N}{{{\tau }_{n}}}{{\mathbf{X}}_{n}}
\label{eq:15},
\end{equation}
where ${{\tau }_{n}}$ is the modal excitation coefficient, which can be expressed as
\begin{equation}
{{\tau }_{n}}=\frac{(\mathbf{X}_{n}^{a}){{}^{\text{T}}}{{\mathbf{F}}^{\text{inc}}}}{1\text{+}j{{\lambda }_{n}}}
\label{eq:19}.
\end{equation}
To reduce the computational costs and simplify the CM solving  procedure, a procedure similar to [4] can be utilized to symmetrize the aforementioned EFIE-PMCHWT equation as following (termed as sEFIE-PMCHWT),

\begin{figure*}[htb]
\begin{equation}
\left[\begin{array}{cccc}
\eta_{1} \mathbf{P}_\mathrm{d, d}^{1}+\eta_{2} \mathbf{P}_\mathrm{d, d}^{2} & j\mathbf{Q}_\mathrm{d, d}^{1}+j\mathbf{Q}_\mathrm{d, d}^{2} & \eta_{1} \mathbf{P}_\mathrm{d, c_{1}}^{1} & \eta_{2} \mathbf{P}_\mathrm{d, c_{2}}^{2} \\
j\mathbf{Q}_\mathrm{d, d}^{1}+j\mathbf{Q}_\mathrm{d, d}^{2} & 1 / \eta_{1} \mathbf{P}_\mathrm{d, d}^{1}+1 / \eta_{2} \mathbf{P}_\mathrm{d, d}^{2} & j\mathbf{Q}_\mathrm{d, c_{1}}^{1} & j\mathbf{Q}_\mathrm{d, c_{2}}^{2} \\
\eta_{1} \mathbf{P}_\mathrm{c_{1}, d}^{1} & j\mathbf{Q}_\mathrm{c_{1}, d}^{1} & \eta_{1} \mathbf{P}_\mathrm{c_{1}, c_{1}}^{1} & 0 \\
\eta_{2} \mathbf{P}_\mathrm{c_{2}, d}^{2} & j\mathbf{Q}_\mathrm{c_{2}, d}^{2} & 0 & \eta_{2} \mathbf{P}_\mathrm{c_{2}, c_{2}}^{2}
\end{array}\right]\left[\begin{array}{c}
\mathbf{J}_\mathrm{d} \\
j\mathbf{M}_\mathrm{d} \\
\mathbf{J}_\mathrm{c_{1}} \\
\mathbf{J}_\mathrm{c_{2}}
\end{array}\right]
=\left[\begin{array}{c}
\mathbf{b}_\mathrm{d}^\text{TE} \\
j\mathbf{b}_\mathrm{d}^\text{TH} \\
\mathbf{b}_\mathrm{c_{1}}^\text{TE} \\
0
\end{array}\right].
\label{eq:20}
\end{equation}
\end{figure*}

Because the exterior part
\begin{equation}
{{\mathbf{Z}}^{\text{ext}}}=\left[ \begin{matrix}
   \eta_{1} \mathbf{P}_\mathrm{d, d}^{1}& j\mathbf{Q}_\mathrm{d, d}^{1}& \eta_{1} \mathbf{P}_\mathrm{d, c_{1}}^{1} & 0 \\
j\mathbf{Q}_\mathrm{d, d}^{1} & 1 / \eta_{1} \mathbf{P}_\mathrm{d, d}^{1}& j\mathbf{Q}_\mathrm{d, c_{1}}^{1} & 0\\
\eta_{1} \mathbf{P}_\mathrm{c_{1}, d}^{1} & j\mathbf{Q}_\mathrm{c_{1}, d}^{1} & \eta_{1} \mathbf{P}_\mathrm{c_{1}, c_{1}}^{1} & 0 \\
0 &0& 0 & 0  \\
\end{matrix} \right]
\label{eq:21}
\end{equation}
is symmetric, it satisfies
\begin{equation}
-\frac{1}{2}{\text {Re}}[ \mathbf{X}_{n}^{\text{H}}\cdot({{\mathbf{Z}}^{\text {ext}}}\cdot{{\mathbf{X}}_{n}}) ] = -\frac{1}{2}[ \mathbf{X}_{n}^{\text{H}}\cdot({\text {Re}}({{\mathbf{Z}}^{\text {ext}}})\cdot{{\mathbf{X}}_{n}}) ].
\label{eq:22}
\end{equation}

Thus, one can define the weighting matrix $\mathbf{W}$ of the symmetric equation with the real part of the operator ${{\mathbf{Z}}^{\text {ext}}}$
\begin{equation}
\mathbf{W}\text{=}\left[ \begin{matrix}
 {\text {Re}}(\eta_{1} \mathbf{P}_\mathrm{d, d}^{1})& {\text {Im}}(-\mathbf{Q}_\mathrm{d, d}^{1})& {\text {Re}}(\eta_{1} \mathbf{P}_\mathrm{d, c_{1}}^{1}) & 0 \\
{\text {Im}}(-\mathbf{Q}_\mathrm{d, d}^{1}) & {\text {Re}}(1 / \eta_{1} \mathbf{P}_\mathrm{d, d}^{1})& {\text {Im}}(-\mathbf{Q}_\mathrm{d, c_{1}}^{1}) & 0\\
{\text {Re}}(\eta_{1} \mathbf{P}_\mathrm{c_{1}, d}^{1}) & {\text {Im}}(-\mathbf{Q}_\mathrm{c_{1}, d}^{1}) & {\text {Re}}(\eta_{1} \mathbf{P}_\mathrm{c_{1}, c_{1}}^{1}) & 0 \\
0 &0& 0 & 0  \\
\end{matrix} \right]
\label{eq:23}
\end{equation}

Compared to the non-symmetric formulation, the symmetric one needs less memory and improves computational efficiency. In particular, in the computation of the induced current, it can avoid constructing the supplementary eigenvalue equation.

\section{Numerical Results}
In the first numerical example, a rectangular metallic film of dimensions 100mm$\times $40mm is investigated. The metallic film covers the entire upper surface of the FR-4 substrate. The thickness and relative dielectric constant of the substrate is 1.55mm and $\varepsilon_r=\mathrm{4.7}$, respectively. The mesh size is 0.3mm. The frequency bands range from 1 to 8 GHz with a frequency step of 50 MHz.

\begin{figure}[!htbp]
\centerline{\includegraphics[width=0.75\columnwidth,draft=false]{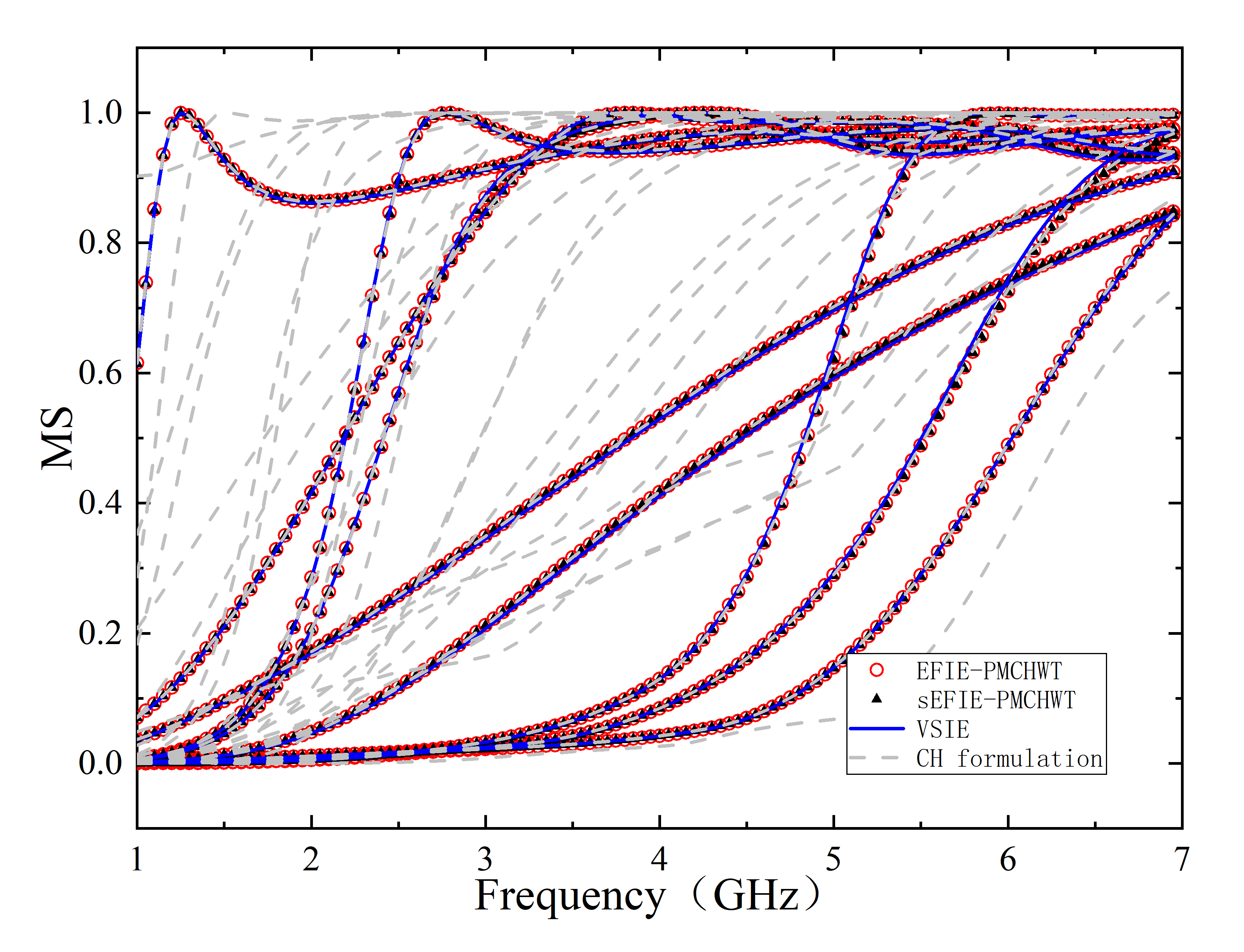}}
\caption{The MS of the CMs about a rectangular patch on FR4 cuboid substrate. Result of CH formulation, EFIE-PMCHWT, and sEFIE-PMCHWT formulations are compared with the results of the VSIE solved by FEKO.}
\label{fig:2}
\end{figure}
\begin{figure}[!htbp]
\centerline{\includegraphics[width=0.8\columnwidth,draft=false]{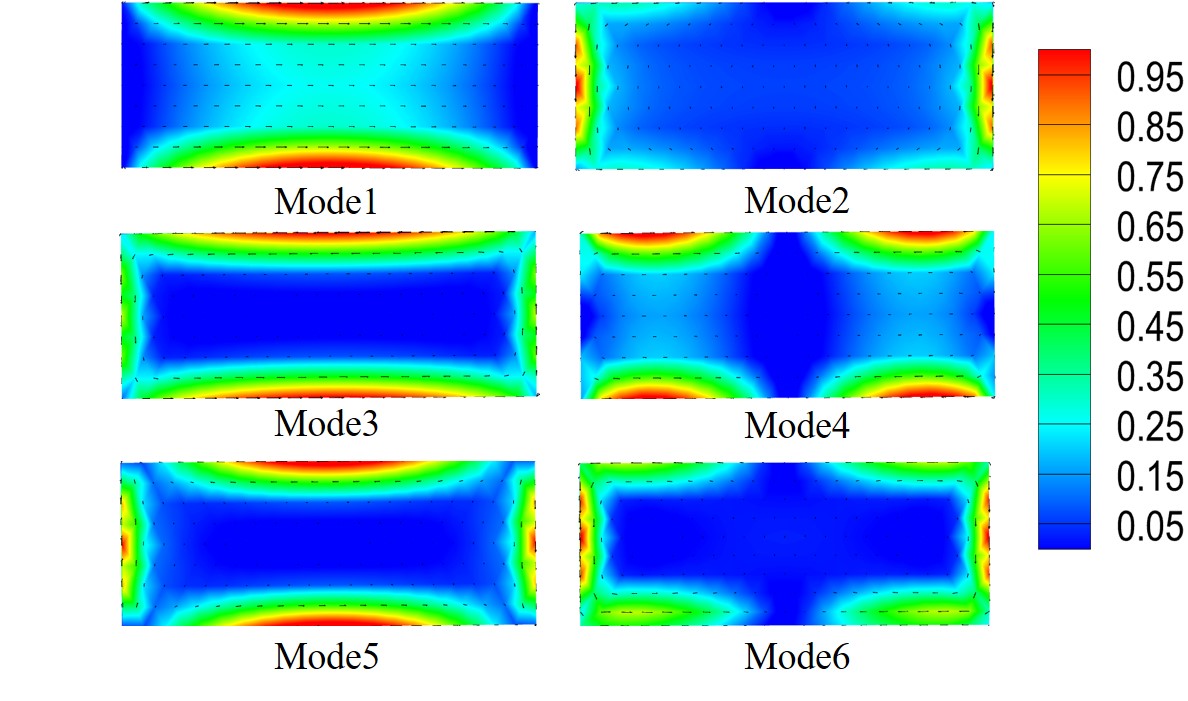}}
\caption{The first 6 electric eigencurrents of the metallic patch at 1.275 GHz computed with the sEFIE-PMCHWT.}
\label{fig:3}
\end{figure}

Fig.~\ref{fig:2} shows the modal significance of first 9 modes that having the biggest modal significance at the lowest frequency. The modal significance is defined as:
\begin{equation}
{\text {MS}=\left| \frac{1}{1+j{{\lambda }_{n}}} \right|}
\label{eq:24}
\end{equation}
The curves with lines represent the results of VSIE solved by commercial software FEKO, which is immune from spurious modes. The curves with hollow circles and solid triangles represent the results of EFIE-PMCHWT and sEFIE-PMCHWT, respectively. The results of these three formulations agree well with each other, showing that the proposed two SIE-based formulations (EFIE-PMCHWT and sEFIE-PMCHWT) for the patch structure can effectively avoid the spurious mode with the same accuracy. In comparison, the results of the CH formulation are also plotted with gray dotted lines. In the CH formulation,  the real part of the impedance matrix is chosen as the right weighting operator. Obviously, it generates a large number of spurious modes. These results validate that the EFIE-PMCHWT and sEFIE-PMCHWT formulations have the same accuracy. In the rest of this section, for simplification, we only chose the results from sEFIE-PMCHWT.

To further validate the proposed method, the electric currents at 1.275GHz is also plotted, which is the resonant frequency of mode 1, corresponding to the same structure shown in Fig.~\ref{fig:2}. Fig.~\ref{fig:3} displays the first six electric currents on the metallic patch surface. The eigen-currents are very similar to the results of \cite{19}, which is solved by VSIE.

%\begin{figure*}[htbp]
%\centerline{\includegraphics[width=1.8\columnwidth,draft=false]{fig3.jpg}}
%\caption{The electric and normalized magnetic currents, left: (a) the currents of 1GHz from MT-DDM-SIE and EFIE-PMCHW,  right: (b) the currents of 3GHz from MT-DDM-SIE. }
%\label{fig:curex3}
%\end{figure*}
%\begin{figure}[!htbp]
%\centerline{\includegraphics[width=0.8\columnwidth,draft=false]{Fig3a.jpg}}
%\caption{The geometry of the circular patch on a rectangle substrate.}
%\label{fig:3a}
%\end{figure}

\begin{figure}[!htbp]
\centerline{\includegraphics[width=0.9\columnwidth,draft=false]{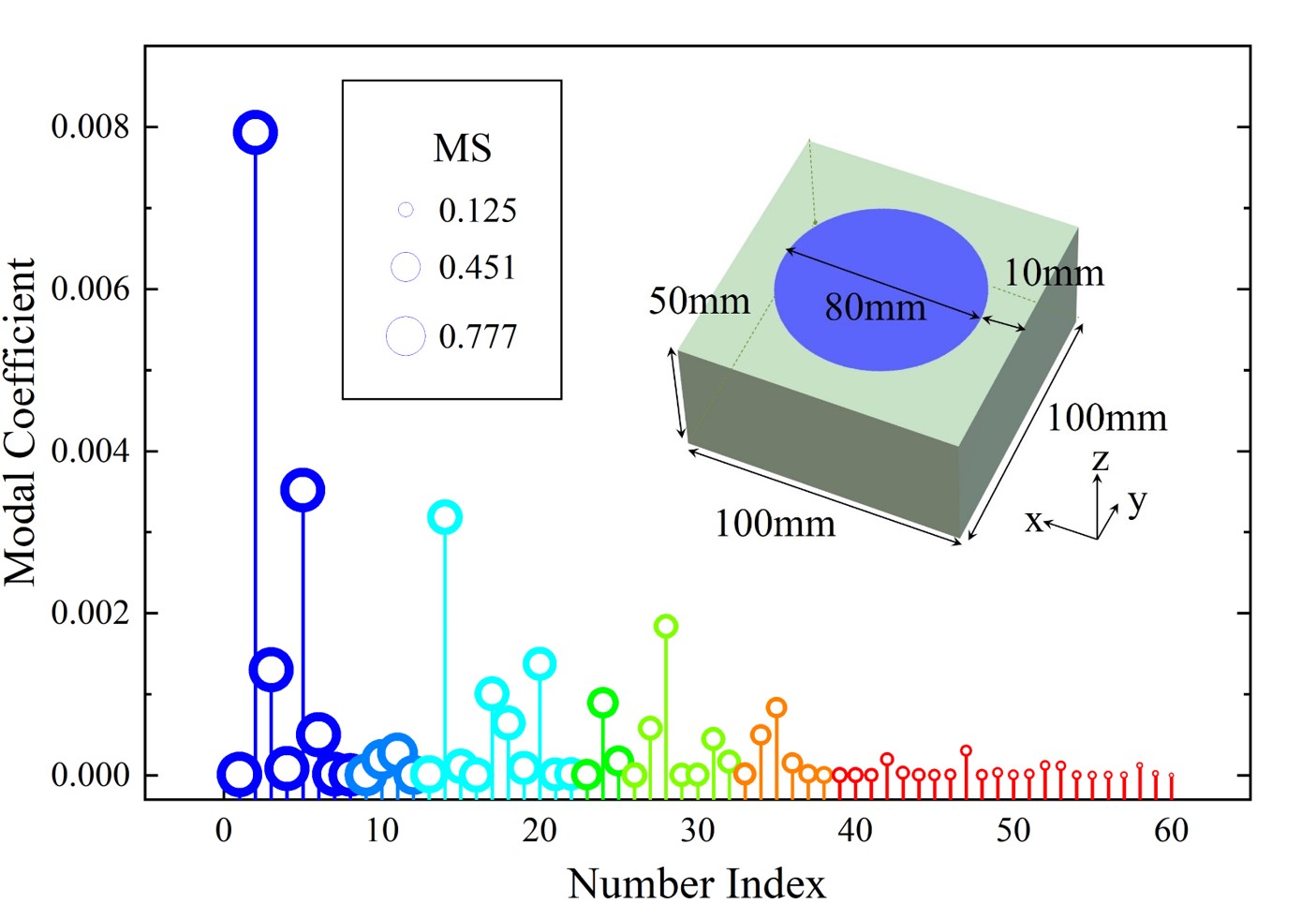}}
\caption{The first 60 modal excitation coefficients based on sEFIE-PMCHWT at 3GHz.}
\label{fig:4}
\end{figure}
\begin{figure}[!htbp]
\centerline{\includegraphics[width=0.9\columnwidth,draft=false]{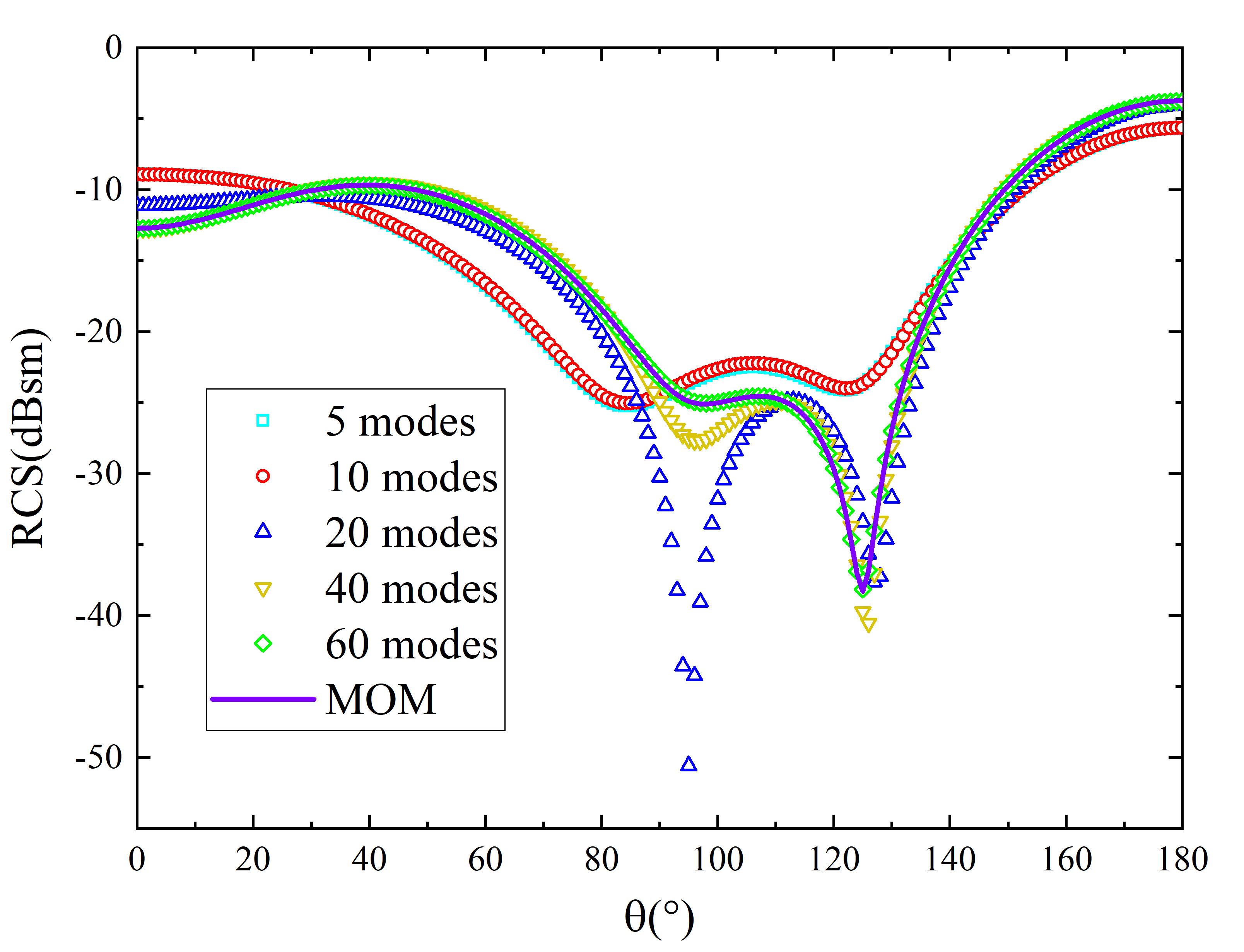}}
\caption{RCSs reconstructed by different number of modes based on sEFIE-PMCHWT at 3GHz for the geometry shown inset of Fig. 4.}
\label{fig:5}
\end{figure}
\begin{figure}[!htbp]
\centerline{\includegraphics[width=0.6\columnwidth,draft=false]{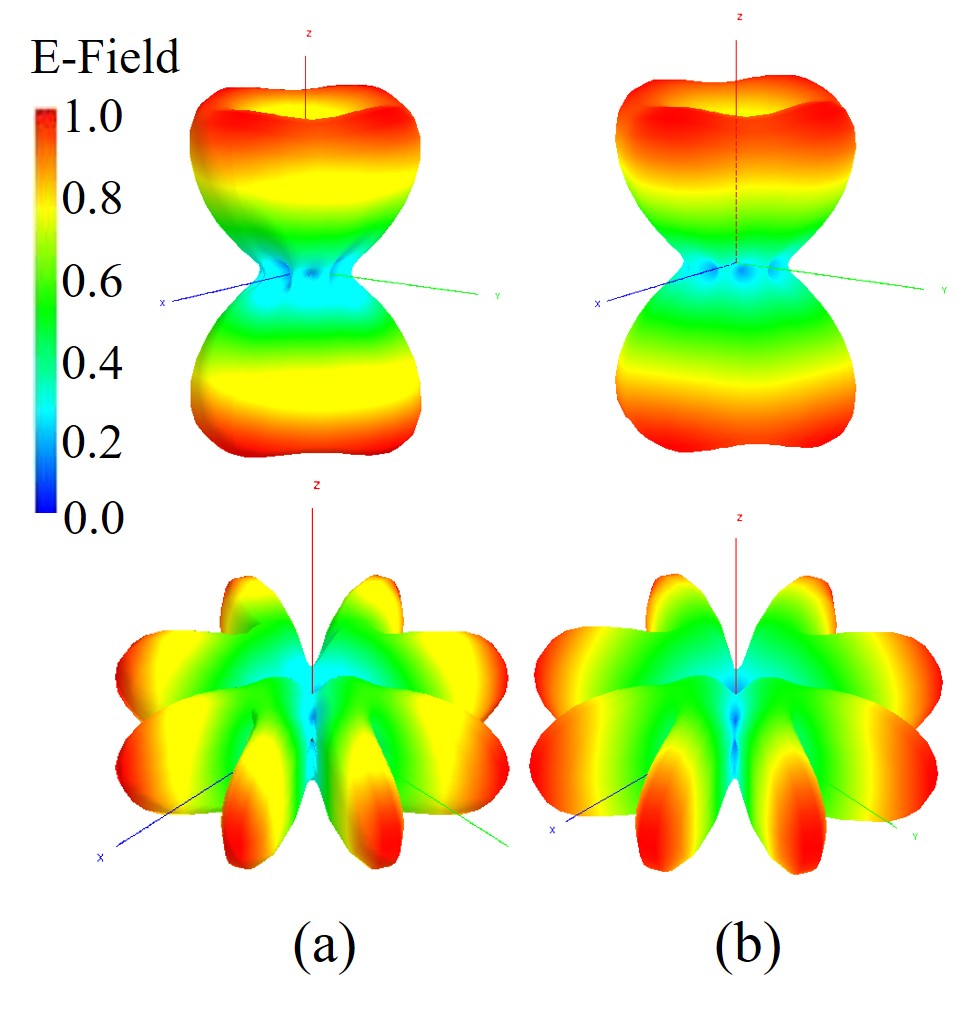}}
\caption{The characteristic fields of mode1 and mode5 computed with (a) VSIE and (b) SIE (sEFIE-PMCHWT) at 3GHz for the geometry shown inset of Fig. 4.}
\label{fig:7}
\end{figure}
In the second numerical example, as shown inset of Fig.~\ref{fig:4}, the radar cross-section (RCS) of a circular patch (with a radius of 40mm) on a thick FR-4 substrate with a size of 100mm$\times $100mm$\times $50mm is investigated. The number of basis functions and modes is 1425 and 2824, respectively. Fig.~\ref{fig:4} plots the first 60 modal excitation coefficients ${{\tau }_{n}}$ at 3 GHz. The mode indexes are sorted from the largest modal significance to the smallest. The figure shows that not all modes can be efficiently excited because the coupling between the modal excitation coefficient and modal significance also depends on some properties of the external source, such as the position, magnitude, phase, and polarization.
Fig.~\ref{fig:5} displays RCSs (the plane wave incidents from -z-axis) reconstructed by 5, 10, 20, 40, and 60 modes (characteristic currents). Evidently, as the number of superimposed modes increases, the RCS converges to the correct result. Good agreement is observed when the reconstructed induced current contains 60 modes.
This property can be used for modeling large-scale finite periodic arrays \cite{22} and analyzing general object's scattering property  \cite{23} with multiple excitations by choosing a small number of CMs with the lowest eigenvalues as entire-domain basis functions.

The characteristic fields of the  proposed SIE-based method (b) are compared with the results from the VSIE formulations (a), in Fig.~\ref{fig:7}. The characteristic fields are produced by the characteristic currents of mode1 and mode5, respectively. A good agreement also can be observed.

\section{Conclusions}
In this letter, an EFIE-PMCHWT based TCM and its symmetrization (sEFIE-PMCHWT) are proposed for patch antenna structures. By defining the radiation-related right weight operator of the generalized eigenvalue equation, spurious modes can be effectively removed. The physical meaning of the right weight operator is provided following Poynting's theorem. Numerical results have shown the accuracy and efficiency of this formulation.

\ifCLASSOPTIONcaptionsoff
  \newpage
\fi

\end{document}